# HIGH LEVEL SOFTWARE FOR 4.8 GHZ LHC SCHOTTKY SYSTEM

J. Cai, E. McCrory, R. Pasquinelli Fermilab, Batavia, IL 60506 USA*, M. Favier, O. R. Jones, T. Lahey CERN Geneva, Switzerland, A. Jansson ESS Lund Sweden

*Abstract*

The performance of the LHC depends critically on the accurate measurements of the betatron tunes. The betatron tune values of each LHC beam may be measured without excitation using a newly installed transverse Schottky monitor[1]. A high-level software package written in Java has been developed for the Schottky system. The software allows end users to monitor and control the Schottky system, and provides them with non-destructive and continuous bunch-by-bunch measurements for the tunes, momentum spreads, chromaticities and emittances of the LHC beams. It has been tested with both proton and lead ion beams at the LHC with very successful results.

## SOFTWARE ARCHITECTURE

The design of the Schottky Software Suite is based on a three-tier architecture, see Figure 1. The upper tier is an application of graphic user interface (GUI), referred to as the Display & Control Application. This GUI is responsible for data display and user interaction.

The middle tier, also called the Monitor Application, runs continuously as a daemon process on a dedicated server. It acquires the FFT trace data from the Schottky front-end, performs data analysis, and logs the results to the logging database. It is also responsible for publishing FFT trace data, measurement results, and machine status through the publish/subscription service of JMS (Java Message Service). The interaction between the user and the Monitor uses RMI (Remote Method Invocation). The user can send messages to Monitor, and through these messages it can control the Schottky front-end electronics, such as the amplifier switches, local oscillator frequencies, bunch gating electronics, oscilloscope multiplexer etc. More than one user can subscribe to the Monitor Application. However, only one user at a time can gain control.

The lower tier controls the Schottky front-end electronics and acquires the data using FESA[2] classes running on a real-time LYNX-OS, communicating to the middle tier through JAPC[3] software components.

## REAL-TIME DATA ANALYSIS

The transverse Schottky spectrum contains information about the tune, momentum spread, chromaticity and emittance, which can be derived from the frequency, amplitudes, and widths of its two betatron sidebands through curve fitting[4]. However, this analysis poses a big challenge due to the coherent signals that are often observed superimposed onto the incoherent Schottky sidebands. These coherent lines are particularly noticeable with LHC proton beams, but are significantly diminished with lead ion beam. The distribution is far from being Gaussian for both proton and lead ion beam. Many empirical fitting functions have been tried. We find the following fitting function works well:

$$F(x) = A_1\, e^{-\frac{1}{2}\left(\frac{x-\mu_1}{\sigma_1}\right)^4} + A_2 e^{-\frac{1}{2}\left(\frac{x-\mu_2}{\sigma_2}\right)^4} + F_0 + S\,x$$

where, $A_1$ and $A_2$ are the amplitude for the first and second sideband, and $\mu_1$, $\mu_2$, $\sigma_1$, and $\sigma 2$ are their corresponding centers and widths. $F_0$ is the overall offset of the spectrum, and S is its slope. To get a satisfactory result, we must remove the coherent signals prior to performing the fits. The algorithm first locates the coherent signal(s), then uses a look-up table to determine the width, and replaces data with linear interpolation in the final fit. The look-up table used was established statistically from a large samples of spectra, which has been seen to be related to the amplitude of the coherent line.

For lead ion beams, where the coherent lines were much smaller, a fixed width is used to remove this signal. The type of particle being accelerated is transmitted to the software via the LHC timing system.

An example of a typical transverse FFT Schottky trace from a proton beam can be seen in Figure 2. It shows the raw data with coherent lines, the reduced data used for the fit with the coherent signals removed, and the final fit results. Notice that the strong spike between the two sidebands, at half the revolution frequency, is also removed automatically.

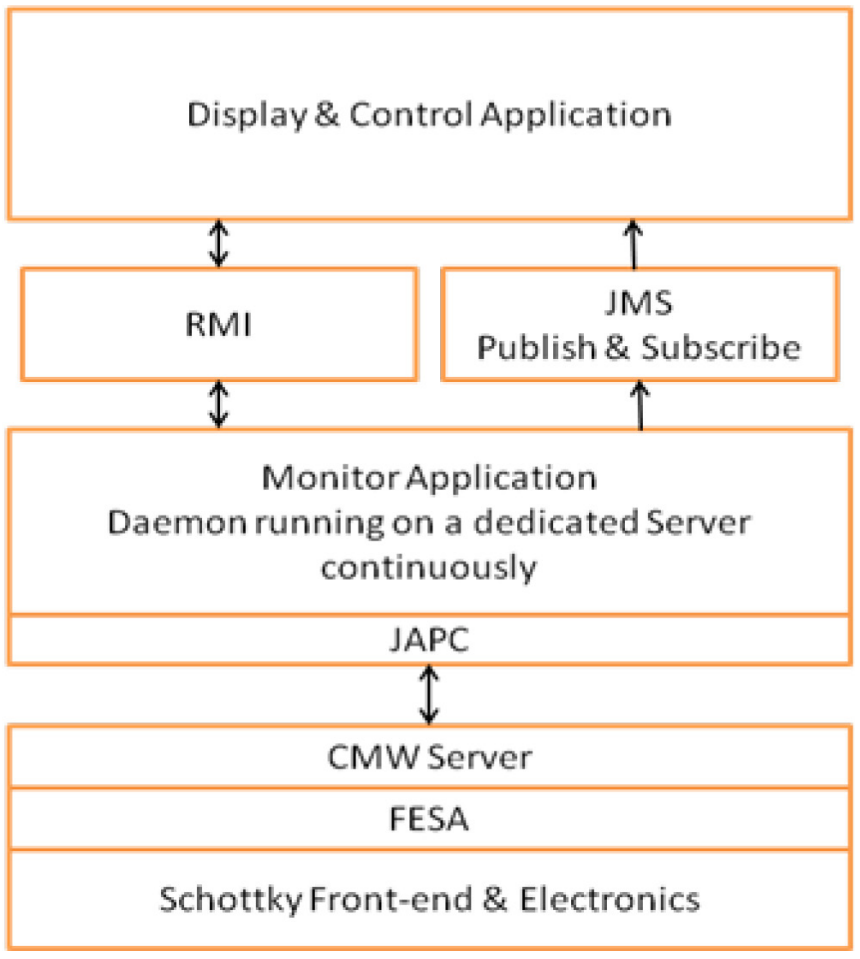

Figure 1. The architecture of the Schottky Software

* Fermilab is operated by Fermi Research Alliance, LLC under Contract No. DE-AC02-07CH11359 with the United States Department of Energy.

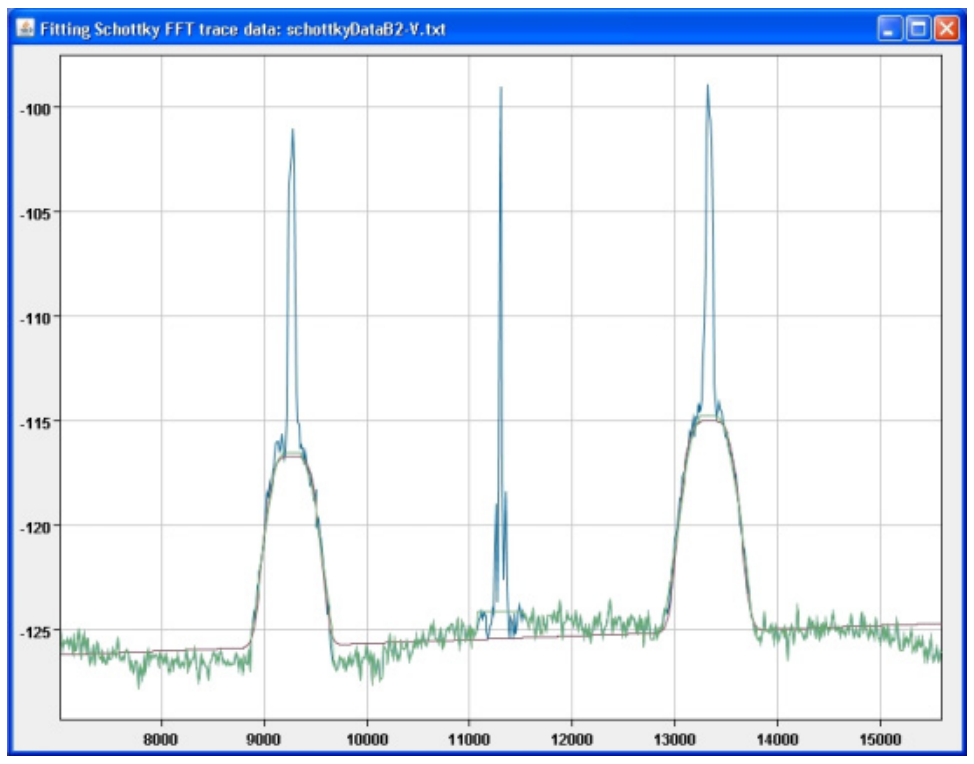

Figure 2. Transverse proton Schottky spectrum (blue), reduced spectrum (green) and the fit (brown).

Fitting is a computationally intensive process. The average fitting time using the MINUIT[5] is about 300 msec for the two planes beam on the current Linux server(cs-ccr-spsea2). These fitting times make publishing the fit results of both planes of both beams possible at rate of 1 Hz.

The tune is calculated from the fitted centers of the two sidebands. Averaging the width of two corresponding sidebands gives the momentum spread, and the chromaticity comes from the relative difference between the widths. The emittance is calculated from the sum of the area of two sidebands, but requires an external calibration factor[6] to give absolute values. This factor can be set by end user from the GUI.

## DISPLAY & CONTROL APPLICATION

The Display & Control Application is the most common way for the end users to display the measurement results and status of the Monitor Application. The end users can also change settings of the Monitor application, including the Schottky front-end and electronics devices from this GUI. Application can be launched from a Web browser using Java Web Start (JWS), or from an LHC console computer. Figure 3 shows a snapshot of the GUI application. The first block on the left panel shows the current Setting Mode, either “Read Only”, or “Read & Set”. If “Read & Set” is chosen, it will also show the name of the user and the number of seconds remaining for this user to have control( between 1 min and 8 hours). During this time only the selected user will be able to make changes to the Schottky system settings.

The Display & Control Application also shows other information, such as the current beam mode (injection, ramp, squeeze, stable beam etc), beam intensity, and the status of the Monitor Application. “Daemon Idle” is the time elapsed since last published data was received, which provides the heartbeat for the Monitor.

There are eleven tabs in the central panel for showing the most important information from the Schottky.

The first tab is a “Data Display” as shown in the Figure 3. This particular example shows FFT trace data taken on 6th December 2010 with lead ion beam, including the fit results. The second tab( Figure 4) shows the time evolution plot for the measured parameters, while the third tab(Figure 5) allows control of the electronics via a circuit diagram and of the local oscillator settings. The circuit diagram is a synoptic display showing the current settings for the amplifier gain and attenuators, and the states of all the control switches used for calibration and selection of signal treatment paths. The end user can change the settings using mouse clicks once he has secured the “Read & Set” privilege. The lower part of the same tab allows control of the frequency and power settings of the two local oscillators per beam.

For nominal operation each LHC ring will contain up to 2808 particle bunches distributed in 3564 possible slots of 25 nanoseconds. In order to allow bunch to bunch measurements and to reduce the effective noise floor, the front-end Schottky electronics contains a fast gate located directly after the first pre-amplifier. The gate tab(Figure 6) allows the position and width of this gate to be defined in a graphical manner in relation to the current filling pattern. The user can change the bunch selection using mouse clicks, or retrieve a saved bunch pattern. The light green color indicates that there is a bunch in that particular slot while the light blue color indicates there is no beam. Darkened colors indicate that the gate is open for that slot. The user can also check the bunch gating using a remote connection to an oscilloscope located with the acquisition and control electronics.

## DISCUSSION AND CONCLUSION

The Schottky Software Suite has been running at the CERN Control Centre for several months, and has proven to be a valuable tool for measurements of tune and chromaticity. In particular, at this time, it is the only LHC diagnostic that can measure the tune of individual bunches. The three-tier design of the software makes continuous measurements and data-logging for offline data analysis possible. Experience from Fermilab Tevatron indicates this is a very useful function. However, there is still a lot of room for further improvements.

The Monitor daemon runs continuously and unattended. Nevertheless, the fitting process occasionally does not return due to a bug in the MINUIT/Java interface. CPU usage soon reaches 100%, and server becomes unresponsive. Because the fitting software is a “canned” product, we have no control over this bug. In response, we have relegated the fitting to a separate Java thread. If it takes too long for fitting, the thread is stopped. We have run the Monitor daemon for several months, and this approach seems to be successful, despite the fact that the stop method of thread in Java has been deprecated[7]. Even though we do not see any problems for our specific application, we are considering new strategies to deal with this issue.

The fitting routines were optimized on a small subset of Schottky FFT spectra, and there will need to be some further fine tuning as the beam intensity gets higher, and condition change. For example, the chromaticity calculation fluctuates because it is calculated from a small width difference of two sidebands. Instead of using the sideband widths as fitting parameters, we are considering using chromaticity and momentum spread directly as fitting parameters. If this can be made to work, we would expect to improve the accuracy of the chromaticity calculation.

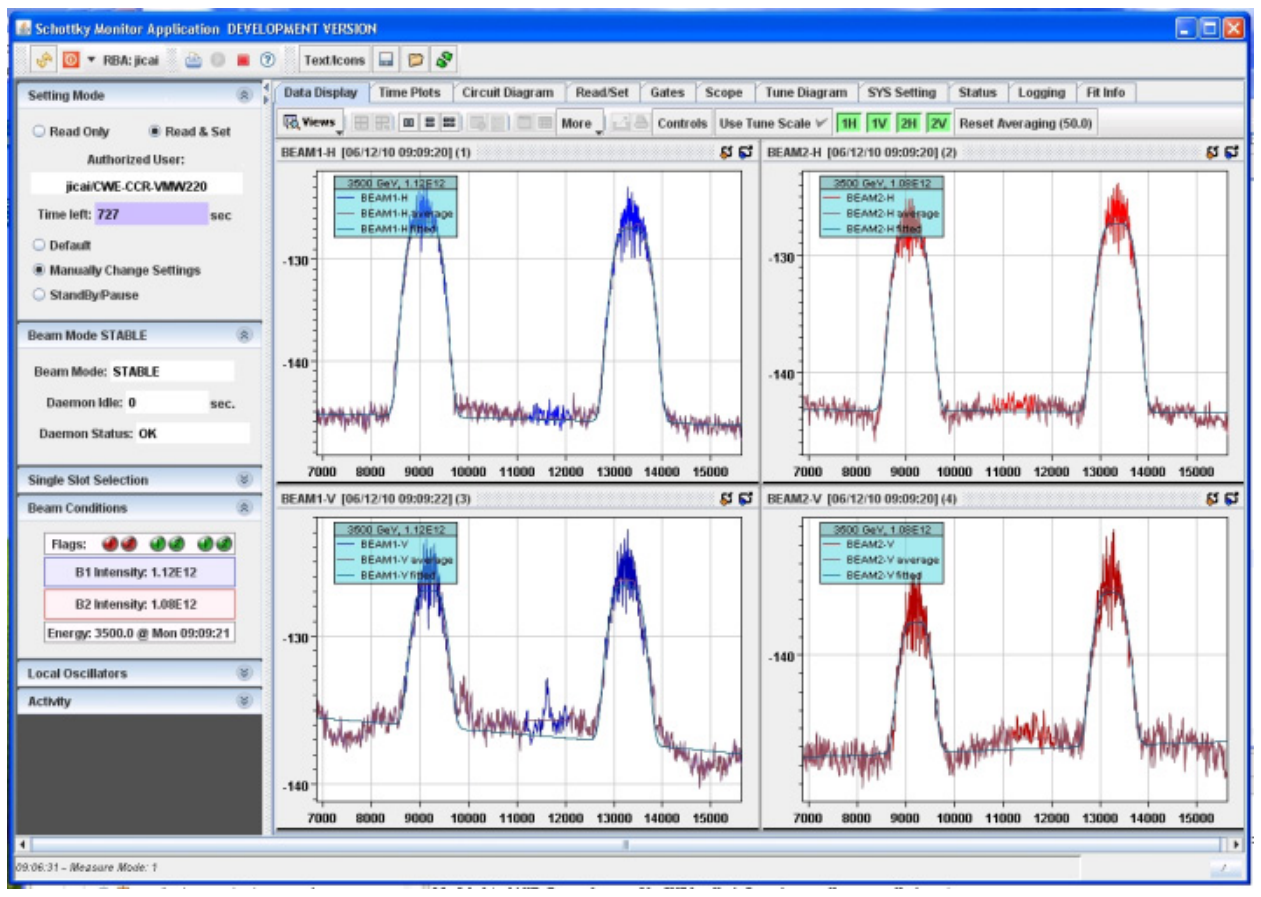

Figure 3. Display and Control Application with Data Display tab shown

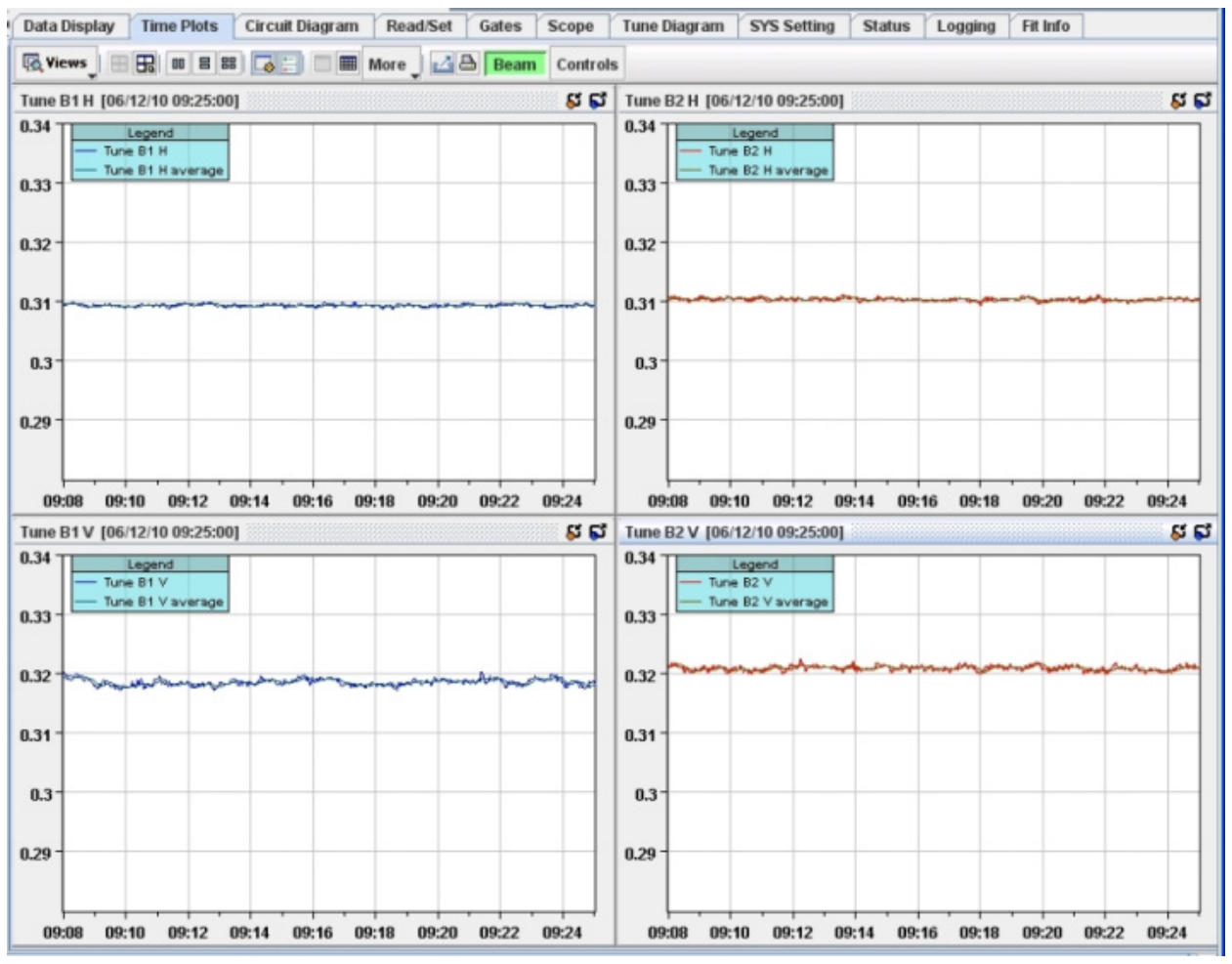

Figure 4. Time evolution plot of tunes for Beam 1 and Beam2 with plane H/V

## ACKNOWLEDGEMENTS

We appreciate the leadership of Dave McGinnis, Suzanne Gysin and Jim Patrick in LAFS team (LHC At Fermilab Software) for many insightful advices and fruitful discussions. Especially, it is Dave McGinnis who provides the proposal on how the software should be designed. We would also like to thank the people in the CERN BE/CO/Applications Section, especially Marek Misiowiec and Katarina Sigerud, for their guidance and assistance with this project.

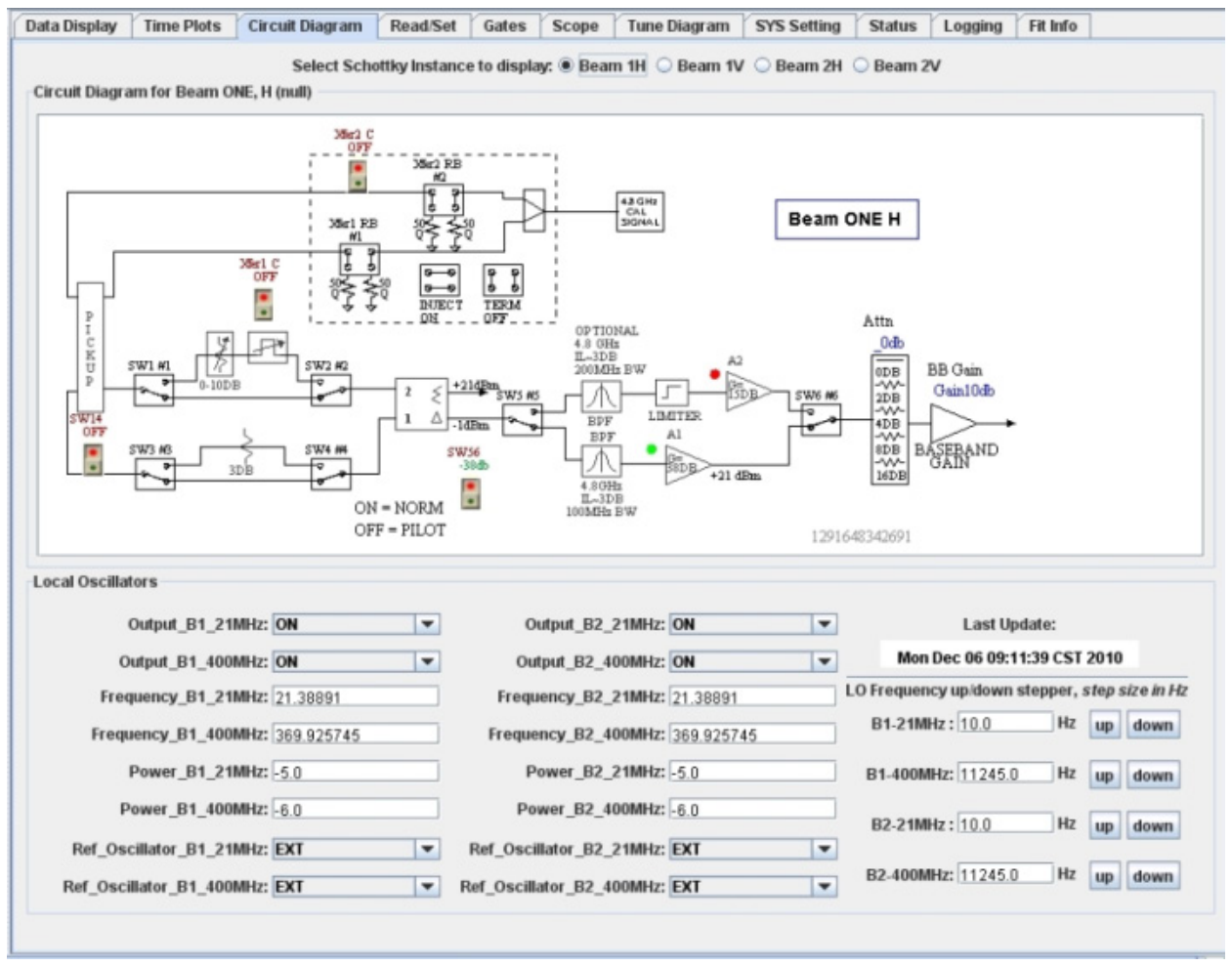

Figure 5. Electronics circuit diagram and local oscillators settings

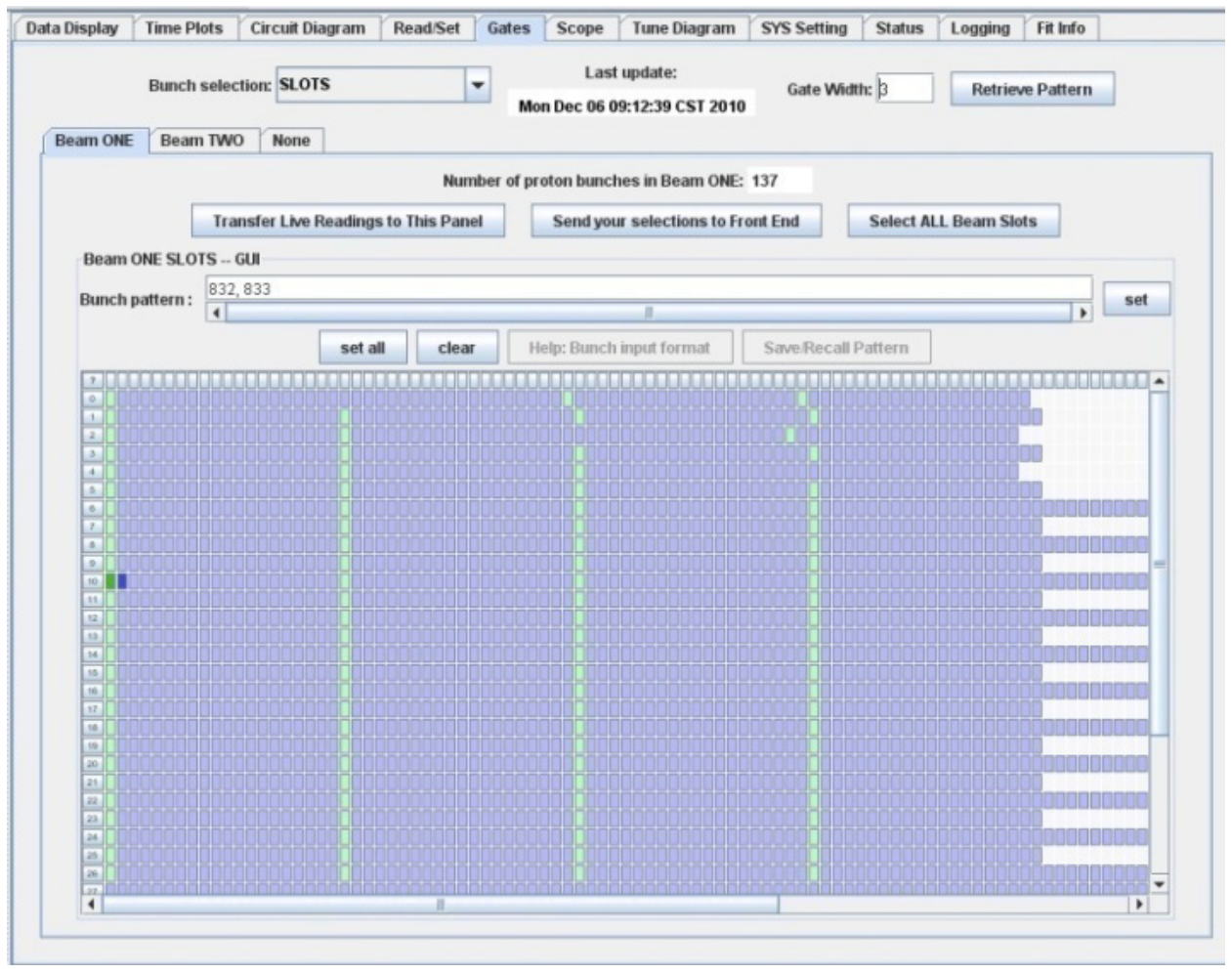

Figure 6. Graphic display for bunch gate settings